\documentclass[12pt]{iopart}
\usepackage{graphicx}

\begin{document}

\title{{A photonic transistor device based on photons and phonons in a cavity electromechanical system}
}
\author{Cheng Jiang$^{1,*}$, and Ka-Di Zhu$^2$}

\address{$^1$School of Physics and Electronic Electrical Engineering, Huaiyin Normal University, 111 West Chang Jiang Road, Huaian 223001, China\\$^2$Key Laboratory of
Artificial Structures and Quantum Control (Ministry of Education),
Department of Physics, Shanghai Jiao Tong University, 800 Dong Chuan
Road, Shanghai 200240, China
} \ead{chengjiang84@yahoo.cn}

\begin{abstract}
We present a scheme for photonic transistors based on photons and
phonons in a cavity electromechanical system, which is consisted of
a superconducting microwave cavity coupled to a nanomechanical
resonator. Control of the propagation of photons is achieved through
the interaction of microwave field (photons) and nanomechanical
vibrations (phonons). By calculating the transmission spectrum of
the signal field, we show that the signal field can be efficiently
attenuated or amplified, depending on the power of a second `gating'
(pump) field. This scheme may be a promising candidate for
single-photon transistors and pave the way for numerous applications
in telecommunication and quantum information technologies.
\end{abstract}

\pacs{42.50.Nn; 85.60.Dw; 85.85.+j; 85.25.-j}
\maketitle
\section{Introduction}
   A photonic transistor is a device where the propagation
of the signal
 photons is controlled by another `gate' photons \cite{Gibbs}. In the past
 decade, such a device has received a lot of interest in view of its
 important applications ranging from optical communication and
 optical quantum computer \cite{Brien} to quantum information processing \cite{Bouwmeester}.
 Schemes based on nanoscale surface plasmons \cite{Chang}, microtoroidal resonators \cite{Hong}, a single-molecule \cite{Hwang}
 and some others \cite{Tominaga,Medhekar,Huang} have been proposed to realize photonic transistors.
 However, practical realization remains challenging because it
 necessitates large nonlinearities and low losses.

 Recently, Weis \emph{et al.} \cite{Weis} reported a novel phenomenon based on the interaction between photons and phonons
 in a cavity optomechanical system, which is called optomechanically induced transparency (OMIT). The transition between the absorptive and transparent
 regime of the probe laser field was modulated by a second control laser field \cite{Weis,Agarwal}.
 In the presence of the control laser, a transparency window for the probe field appears due to the optomechanical
interference effect when the beating of the two laser fields is
resonant with the vibration frequency of the mechanical resonator.
At the same time, electromagnetically induced transparency (EIT) and
slow light with optomechanics have been also experimentally realized
\cite{Safavi}. As a conterpart of cavity optomechanical system,
circuit cavity electromechanical system, usually consisted of a
superconducting microwave cavity and a nanomechanical resonator, has
been under extensive investigation in recent years
\cite{Regal,Vitali,Woolley,Rocheleau}. In particular,
strong-coupling regime was reached recently \cite{Teufel}, which
paves the way for ground-state cooling of the nanomechanical
resonator. Based on these achievements, and we note that they mainly
consider the situation where the cavity is driven on the red
sideband, in the present paper, we deal with a different case that
the cavity is driven on its blue sideband in a coupled
superconducting microwave cavity-nanomechanical resonator system
\cite{Regal,Vitali,Woolley,Rocheleau,Teufel}. Theoretical study
shows that the transmitted signal (`source') field can be attenuated
or amplified, depending on the power of the pump (`gate') field that
controls the number of photons in the cavity. Therefore, such a
system can be employed to serve as a photonic transistor and may be
realized under the existing experimental conditions
\cite{Rocheleau,Teufel}. Importantly, the photonic transistor proposed here
does not require the large nonlinearities and is compatible with
chip-scale processing.
\section{Model and Theory}
Our cavity electromechanical system, composed of a nanomechanical
resonator capacitively coupled to a superconducting microwave cavity
denoted by equivalent inductance $L$ and equivalent capacitance $C$,
is sketched in Fig. 1. A strong pump (`gate') field with frequency
$\omega_{p}$ and a weak signal (`source') field with frequency
$\omega_{s}$ are applied to the microwave cavity simultaneously. The
beating of the two fields causes the nanomechanical resonator to
vibrate which can change the capacitance of the microwave cavity and
thus its resonance frequency. The coupling capacitance can be
approximated by $C_0(x)=C_0(1-x/d)$, where $C_{0}$\ represents an
equilibrium capacitance, \textit{d} is the equilibrium
nanoresonator-cavity separation, and $x$ is the displacement of the
nanomechanical resonator from its equilibrium position. Therefore,
the coupled cavity has an equivalent capacitance
$C_\Sigma=C+C_0(x)$, such that resonance frequency of the microwave
cavity is $\omega_{c}=1/\sqrt{LC_\Sigma}$. In a rotating frame at
the pump frequency $\omega_{p}$, the system Hamiltonian reads as
follows \cite{Regal,Woolley}: {\setlength\arraycolsep{1pt}
\begin{equation}
\fl H=\hbar\Delta_pa^\dagger a+\hbar\omega_nb^\dagger b-\hbar\lambda
a^\dagger
a(b^\dagger+b)+i\hbar(E_pa^\dagger-E_p^*a)+i\hbar(E_sa^\dagger
e^{-i\delta t}-E_s^*ae^{i\delta t}).
\end{equation}
The first term is the energy of the microwave cavity, where
$a^\dagger$ $(a)$ is the creation (annihilation) operator of the
microwave cavity and $\Delta_p=\omega_c-\omega_p$ is the cavity-pump
field detuning. The second term gives the energy of the
nanomechanical resonator with creation (annihilation) operator
$b^\dagger$ $(b)$, resonance frequency $\omega_{n}$ and effective
mass $m$. The third term corresponds to the capacitive coupling
between the microwave cavity and the nanomechanical resonator, where
$\lambda=g\delta x_{zp}$ is the coupling strength between the cavity
and the resonator, $g=\frac{\partial\omega_c}{\partial x}$ is the
effect of the displacement $x=(b^\dagger+b)\delta x_{zp}$ on the
perturbed cavity resonance frequency, and $\delta
x_{zp}=\sqrt{\frac{\hbar}{2\omega_nm}}$ is the zero-point motion of
the nanomechanical resonator. The last two terms represent the
interaction between the cavity field and the two rf fields with
frequency $\omega_p$ and $\omega_s$. $\delta=\omega_s-\omega_p$ is
the signal-pump detuning. $E_{p}$ and $E_{s}$ are, respectively,
amplitudes of the pump field and the signal field, and they are
defined by $\left\vert
E_p\right\vert=\sqrt{2P_p\kappa/\hbar\omega_p}$ and $\left\vert
E_s\right\vert=\sqrt{2P_s\kappa/\hbar\omega_s}$, where $P_{p}$ is
the pump power, $P_{s}$ is the power of the signal field, and
$\kappa$ is the decay rate of the cavity.

Applying the Heisenberg equations of motion for operators $a$ and $Q$ and introducing
the corresponding damping and noise terms \cite{Dobrindt,Genes}, we derive the quantum Langevin equations as follows:
\begin{eqnarray}
&\dot{a}=-(i\Delta_p+\kappa)a+i\lambda aQ+E_p+E_s e^{-i\delta t}+\sqrt{2\kappa}a_{in},\\
&\ddot{Q}+\gamma_n \dot{Q}+\omega_n^2Q=2\omega_n\lambda a^\dagger a+\xi,
\end{eqnarray}
where we have set $Q=b^\dagger+b$ be the resonator amplitude. The cavity mode decays at the rate $\kappa$ and is affected by the
input vacuum noise operator $a_{in}$ with zero mean value, which obeys
the correlation function in the time domain,
\begin{eqnarray}
&\langle\delta a_{in}(t)\delta a_{in}^\dagger(t^\prime)\rangle=\delta(t-t^\prime),\\
&\langle\delta a_{in}(t)\delta a_{in}(t^\prime)\rangle=\langle\delta a_{in}^\dagger(t)\delta a_{in}(t^\prime)\rangle=0.
\end{eqnarray}
The mechanical mode is affected by a vicious force with damping rate $\gamma_n$ and by a Brownian stochastic force with zero mean value $\xi$ that has the following correlation function \cite{Giovannetti}
\begin{eqnarray}
\langle\xi(t)\xi(t^\prime)\rangle=\frac{\gamma_n}{\omega_n}\int \frac{d\omega}{2\pi}\omega e^{-i\omega (t-t^\prime)}\left[1+\mathrm{coth}\left(\frac{\hbar\omega}{2k_BT}\right)\right],
\end{eqnarray}
where $k_B$ is the Boltzmann constant and $T$ is the temperature of the reservoir of the mechanical resonator. Following standard methods from quantum optics,
we derive the steady-state solution to Eqs. (2) and (3) by setting all the time derivatives to zero. They are given by
\begin{eqnarray}
a_s=\frac{E_p}{\kappa+i(\Delta_p-\lambda Q_s)}, Q_s=\frac{2\lambda\left\vert a_s\right\vert^2}{\omega_n}.
\end{eqnarray}
To go beyond weak coupling, we can always rewrite each Heisenberg operator as the sum of its steady-state mean value and a small fluctuation with zero mean value,
\begin{eqnarray}
a=a_s+\delta a, Q=Q_s+\delta Q.
\end{eqnarray}Inserting this equation into the Langevin equations Eqs. (2)-(3) and assuming $\left\vert a_s\right\vert\gg1,$
one can safely neglect the nonlinear terms $\delta a^\dagger\delta a$ and $\delta a\delta Q$ and get the linearized Langevin equations \cite{Weis,Dobrindt,Marquardt},
\begin{eqnarray}
&\delta\dot{a}=-(\kappa+i\Delta_p)\delta a+i\lambda Q_s\delta a+i\lambda a_s\delta Q+E_s e^{-i\delta t}+\sqrt{2\kappa}a_{in},\\
&\delta\ddot{Q}+\gamma_n\delta\dot{Q}+\omega_n^2\delta
Q=2\omega_n\lambda a_s(\delta a+\delta a^\dagger)+\xi.
\end{eqnarray}
Such linearized Langevin equations have been used in investigating
optomechanically induced transparency \cite{Weis}, parametric
normal-mode splitting \cite{Dobrindt}, and sideband cooling of
mechanical motion \cite{Marquardt} in optomechanical systems, where
the strong-coupling is required. In the circuit cavity
electromechanical system we study here, strong-coupling regime has
been achieved recently, where the cooperativity $C\approx2000,000$
\cite{Teufel}, larger than those previously achieved in
optomechanical systems \cite{Weis,Groblacher}. Here,
$C=4G^2/\gamma_n\kappa$ ($G=\lambda\sqrt n_p$ is the effective
coupling strength) is an equivalent opto- or electro-mechanical
cooperativity parameter. In this situation, the eigenmodes of the
driven system are hybrids of the original mechanical mode and the
cavity mode. The coupled system shows the normal-mode splitting, a
phenomenon well-known to both classical and quantum physics.
Theoretical transmission spectrum obtained through the linearized
Langevin equations is in good agreement with the experiment result
\cite{Teufel}. In the following, since the drives are weak, but
classical coherent fields, we will identify all operators with their
expectation values, and drop the quantum and thermal noise terms
\cite{Weis}. Then the linearized Langevin equations can be written
as:
\begin{eqnarray}
&\left\langle\delta\dot{a}\right\rangle=-(\kappa+i\Delta_p)\left\langle\delta a\right\rangle+i\lambda Q_s\left\langle\delta a\right\rangle+i\lambda a_s\left\langle\delta Q\right\rangle+E_s e^{-i\delta t},\\
&\langle\delta\ddot{Q}\rangle+\gamma_n\langle\delta\dot{Q}\rangle+\omega_n^2\langle\delta Q\rangle=2\omega_n\lambda a_s(\langle\delta a\rangle+\langle\delta a^\dagger\rangle).
\end{eqnarray}
In order to solve Eqs. (11) and (12), we make the ansatz \cite{Boyd,Kippenberg1} $\langle\delta a\rangle=a_+e^{-i\delta
t}+a_-e^{i\delta t}$, and $\langle\delta Q\rangle=Q_+e^{-i\delta
t}+Q_-e^{i\delta t}$. Upon substituting the above ansatz into Eqs. (11) and (12), we derive the following solution

\begin{eqnarray}
a_+=\frac{i(\delta+\Delta_p)-(\kappa+\theta)}{(\delta+i\kappa)^2+(\theta-i\Delta_p)^2+\beta}
E_s,
\end{eqnarray}
where $\eta=\frac{\omega_n^2}{\omega_n^2-\delta^2-i\gamma_n\delta}$,
$\alpha=\frac{2\lambda^2}{\omega_n^2}$,
$\beta=\alpha^2\eta^2\omega_n^2n_p^2$,
$\theta=i\alpha\omega_nn_p(\eta+1)$, and $n_p=\left\vert
a_s\right\vert^2$. Here $n_p$, approximately equal to the number of
pump photons in the cavity, is determined by the following equation
\begin{eqnarray}
n_p\left[\kappa^2+\left(\Delta_p-\omega_{n}\alpha
n_p\right)^2\right]=\left\vert E_p\right\vert^2.
\end{eqnarray}
This form of cubic equation is characteristic of optical
multistability \cite{Gupta,Kanamoto}.

The output field can be obtained by employing the standard
input-output theory \cite{Gardiner}
$a_{out}(t)=a_{in}(t)-\sqrt{2\kappa}a(t)$, where $a_{out}(t)$ is the
output field operator, we have
\begin{eqnarray}
\fl \left\langle a_{out}(t)\right\rangle=(E_p-\sqrt{2\kappa}a_s)
e^{-i\omega_p t}+(E_s-\sqrt{2\kappa}a_+)
e^{-i(\delta+\omega_p)t}-\sqrt{2\kappa}a_- e^{i(\delta-\omega_p)t}.
\end{eqnarray}
The transmission of the signal field, defined by the ratio of the
output and input field amplitudes at the signal frequency, is then
given by
\begin{eqnarray}
t_{p}&=&\frac{E_s-\sqrt{2\kappa}a_+}{E_s}=1-\sqrt{2\kappa}a_+/E_s.
\end{eqnarray}
\section{Numerical results and discussion}
In what follows, we choose a realistic cavity electromechanical
system to calculate the transmission spectrum of the signal field.
The parameters used in the numerical simulation are
\cite{Rocheleau}: $\omega_{c}=2\pi\times7.5$ GHz,
$\omega_{n}=2\pi\times6.3$ MHz, $\kappa=2\pi\times600$ kHz,
$\lambda=250$ Hz, and $Q_{n}=10^{6},$ where $Q_{n}$ is the quality
factor of the nanomechanical resonator, and the damping rate $\gamma
_{n}$ is given by $\frac{\omega _{n}}{Q_n}.$ We can see that
$\omega_{n}>\kappa$, therefore the system operates in the
resolved-sideband regime also termed good-cavity limit. When the
cavity is driven on its red sideband, i.e. $\Delta_p=\omega_n$, the
analogy of electromagnetically induced transparency (EIT) could
appear, which has been extensively discussed
\cite{Weis,Agarwal,Safavi,Teufel}. If we choose $\Delta_p=\omega_n$, we can
also obtain the similar EIT effect, as shown in Fig. 2(a). When the pump power increases from zero, the transmission of the signal field at signal-cavity detuning $\Delta_s=0$ increases to unity gradually, and then the transparency window is broadened by increasing the pump power further. Our result is good agreement with that in Ref. [17]. Here, we mainly consider the
situation where the cavity is driven on its blue sideband, i.e.,
$\Delta_p=-\omega_n$. Under blue-detuned pumping, the effective
interaction Hamiltonian for the cavity field and the mechanical
phonon mode becomes one of the parametric amplification,
$H_{int}=\hbar G(a^\dagger b^\dagger+ab)$, where $G=\lambda\sqrt
n_p$ is the effective coupling strength. Fig. 2(b) displays a series of
transmission spectra of the signal field as a function of the
signal-cavity detuning ($\Delta_s=\omega_s-\omega_c$) for various
pump powers. When the pump field is off, i.e., $P_p=0$, the transmission spectrum of the signal field shows the usual
Lorentzian line shape of the bare cavity. However, as the pump power
is raised ($P_p=0.3$pW and $P_p=0.5$pW), we can see that the
transmission is attenuated around the signal-cavity detuning
$\Delta_s=0$ compared to the situation where the pump field is off, a result of the
increased feeding of photons into the cavity. If the pump power is
increased further, the system switches from electromagnetically
induced absorption (EIA) \cite{Lezama} to parametric amplification
(PA) \cite{Mollow}, leading to signal amplification ($P_p=0.6, 0.8, 0.9$pW).
 When the pump power equals to 0.9pW, the transmitted signal field can be amplified greatly. Similar to Fig. 2(a), at the low pump power, the normal-mode splitting is not apparent even if the coupled system has entered the strong-coupling regime. When the pump power is large enough ($P_p=10$nW), normal-mode splitting can be easily observed. In the strong-coupling regime, the eigenmodes of the coupled system are hybrids of the original cavity modes characterized by $n_p$ cavity photons and mechanical modes represented by $n_m$ mechanical quanta. Fig. 2(c) is the level diagram of the driven, coupled system. When the cavity is driven on its red sideband, corresponding to the transition between $\left\vert n_p,n_m+1\right\rangle$ and $\left\vert n_p+1,n_m\right\rangle$, the system exhibits the mechanical analogue of EIT. However, the blue-detuned pump field induces a transition between $\left\vert n_p,n_m\right\rangle$ and $\left\vert n_p+1,n_m+1\right\rangle$, which can efficiently amplify the signal field at the the cavity resonance. Fig. 2(b) demonstrates
that such a circuit cavity electromechanical system can indeed act
as a photonic transistor, where the pump (`gate') field regulates
the flow of the signal (`source') field by controlling the number of
photons in the cavity. We mainly use the narrow region around
signal-cavity detuning $\Delta_s\approx0$ to attenuate or amplify
the signal field. The physical origin of this phenomenon comes from
the radiation pressure force oscillating at the beat frequency
between the pump field and the signal field, which induces the
vibration of the nanomechanical resonator. When the beat frequency
is resonant with the mechanical resonance frequency $\omega_n$. The
frequency of the pump field $\omega_p$ is downshifted to the Stokes
frequency $\omega_p-\omega_n$, which is degenerate with the signal
field. Constructive interference between the Stokes field and the
signal field amplifies the weak signal field. Similar amplification
of a signal due to radiation pressure backaction in a detuned cavity
optomechanical system was recently demonstrated by Verlot \emph{et
al} \cite{Verlot}. Note that the phenomenon of parametric
oscillation instability can occur at some pump power threshold when
the the Stokes field coincides with the cavity resonance, which has
been predicted by Braginsky \cite{Braginsky} and demonstrated for
the first time at Caltech \cite{Kippenberg}. To better demonstrate
the transistor action of the system, we plot the transmission of the
signal field as a function of the signal-cavity detuning in Fig.
3(a) for $P_p=0.8$pW, $P_p=0.9$pW and $P_p=1.0$pW, respectively. It
is clearly seen that transmission is greatly enhanced around the
signal-cavity detuning $\Delta_s=0$ at the higher pump power. Fig.
3(b) summarizes the transistor characteristic curve by plotting the gain of the transmitted signal field as a function of the pump power when the
two-photon detuning $\delta=\omega_n$. The transmitted signal field
can be amplified when the pump power increases above a critical value. Similar results have been obtained experimentally recently \cite{Massel}.
Therefore, the transmitted signal field can be attenuated or
amplified in this coupled system under the control of the strong
pump field when the cavity-pump detuning $\Delta_p=-\omega_n$.
\section{Conclusion}
In conclusion, we have demonstrated that the coupled nanomechanical
resonator-microwave cavity system can serve as a photonic transistor
when the cavity is pumped on its blue sideband. The transmitted
signal field can be amplified greatly, depending on the power of the
pump field. The photonic transistor is based on the interaction of
photons and phonons, where the pump field (photons) can be converted
into mechanical vibrations (phonons). Quantum interference effect
between the generated Stokes field and the signal field is
responsible for the transistor action of the coupled system. This
scheme proposed here can be achievable immediately in current
experiments \cite{Teufel}.

\ack The authors gratefully acknowledge support
from National Natural Science Foundation of China (No.10774101 and
No.10974133).

\newpage
\centerline{\large{\bf Figure Captions}}

Figure 1 Schematic of a nanomechanical resonator capacitively
coupled to a microwave cavity denoted by equivalent inductance $L$
and equivalent capacitance $C$ in the presence of a strong pump
(`gate') field $\omega_{p}$ and a weak signal (`source') field
$\omega_{s}$. The transmitted signal field can be probed using a
low-noise high-electron-mobility-transistor (HEMT) microwave
amplifier.

Figure 2 The normalized magnitude of the cavity transmission $|
t_p|^2$ as a function of signal-cavity detuning
$\Delta_{s}=\omega_s-\omega_c$ for various pump powers with (a) $\Delta_p=\omega_n$ and (b) $\Delta_p=-\omega_n$, respectively. (c) Level diagram of the coupled system under red-detuned pumping ($\Delta_p=\omega_n$) and blue-detuned pumping ($\Delta_p=-\omega_n$), respectively. The signal field probes the transition in which the mechanical occupation is unchanged.
Other parameters used are $\omega_{c}=2\pi\times7.5$ GHz, $\kappa =2\pi\times 600$ kHz,
$\lambda=250$ Hz, $\gamma_{n}$=40 Hz, and $\omega_{n}=2\pi\times6.3$
MHz.

Figure 3 (a) Amplification of the signal field around the region
$\Delta_s=0$ for three different pump powers with $\Delta_p=-\omega_n$. (b) The photonic
transistor characteristic curve by plotting the gain of the transmitted
signal field with respect to the pump power when the signal field
is resonant with the cavity resonance frequency.

\clearpage
\begin{figure}
\includegraphics[width=12cm]{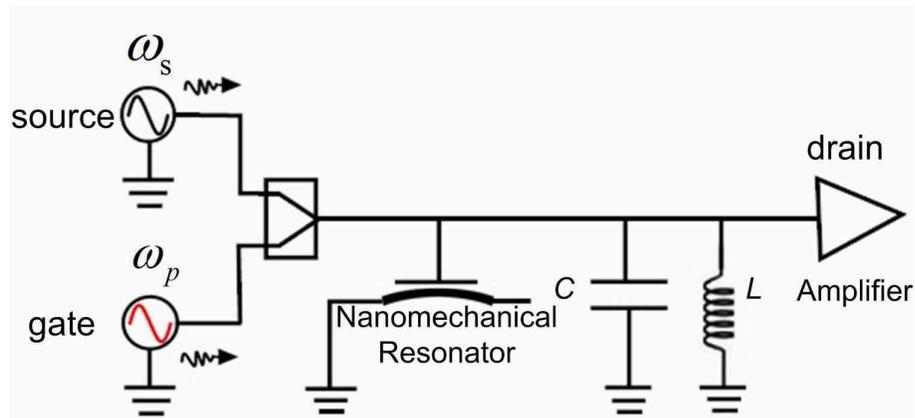}
\caption{Schematic of a nanomechanical resonator capacitively
coupled to a microwave cavity denoted by equivalent inductance $L$
and equivalent capacitance $C$ in the presence of a strong pump
(`gate') field $\omega_{p}$ and a weak signal (`source') field
$\omega_{s}$. The transmitted signal field can be probed using a
low-noise high-electron-mobility-transistor (HEMT) microwave
amplifier.}
\end{figure}

\clearpage
\begin{figure}
\includegraphics[width=12cm]{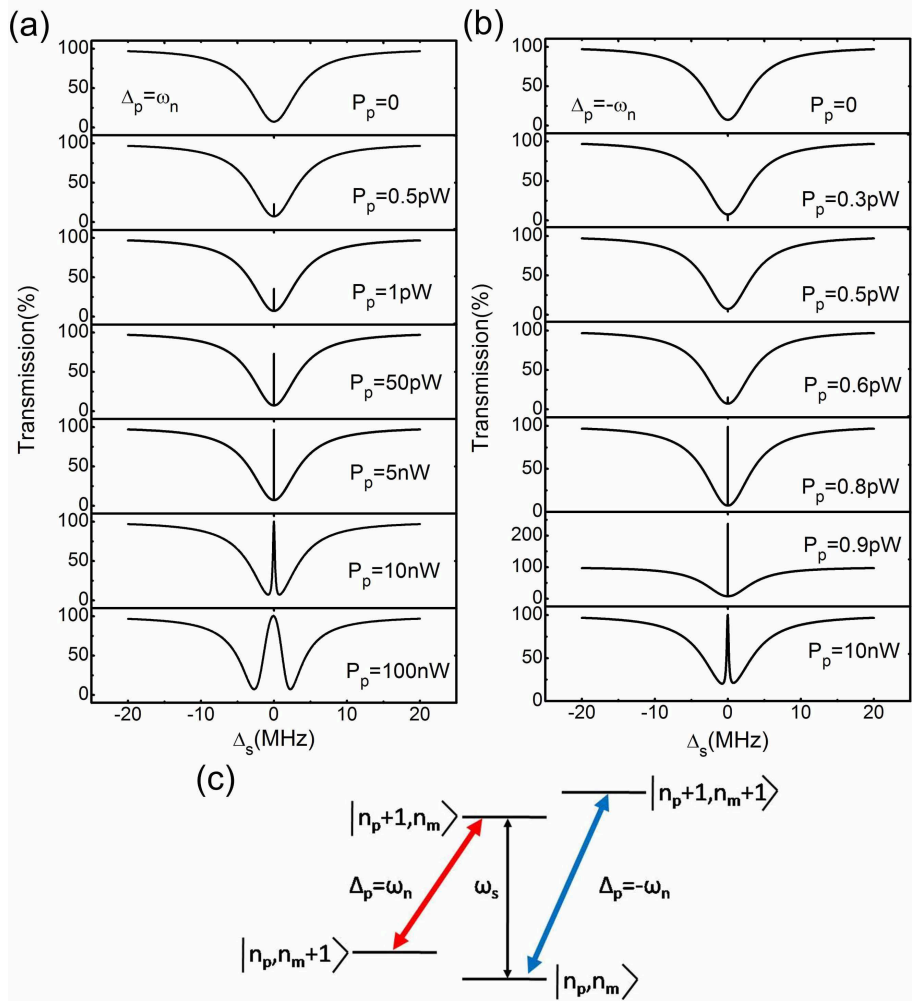}
\caption{The normalized magnitude of the cavity transmission $|
t_p|^2$ as a function of signal-cavity detuning
$\Delta_{s}=\omega_s-\omega_c$ for various pump powers with (a) $\Delta_p=\omega_n$ and (b) $\Delta_p=-\omega_n$, respectively. (c) Level diagram of the coupled system under red-detuned pumping ($\Delta_p=\omega_n$) and blue-detuned pumping ($\Delta_p=-\omega_n$), respectively. The signal field probes the transition in which the mechanical occupation is unchanged.
Other parameters used are
$\omega_{c}=2\pi\times7.5$ GHz, $\kappa =2\pi\times 600$ kHz,
$\lambda=250$ Hz, $\gamma_{n}$=40 Hz, and $\omega_{n}=2\pi\times6.3$
MHz.}
\end{figure}

\clearpage
\begin{figure}[!t]
\centering
\includegraphics[width=12cm]{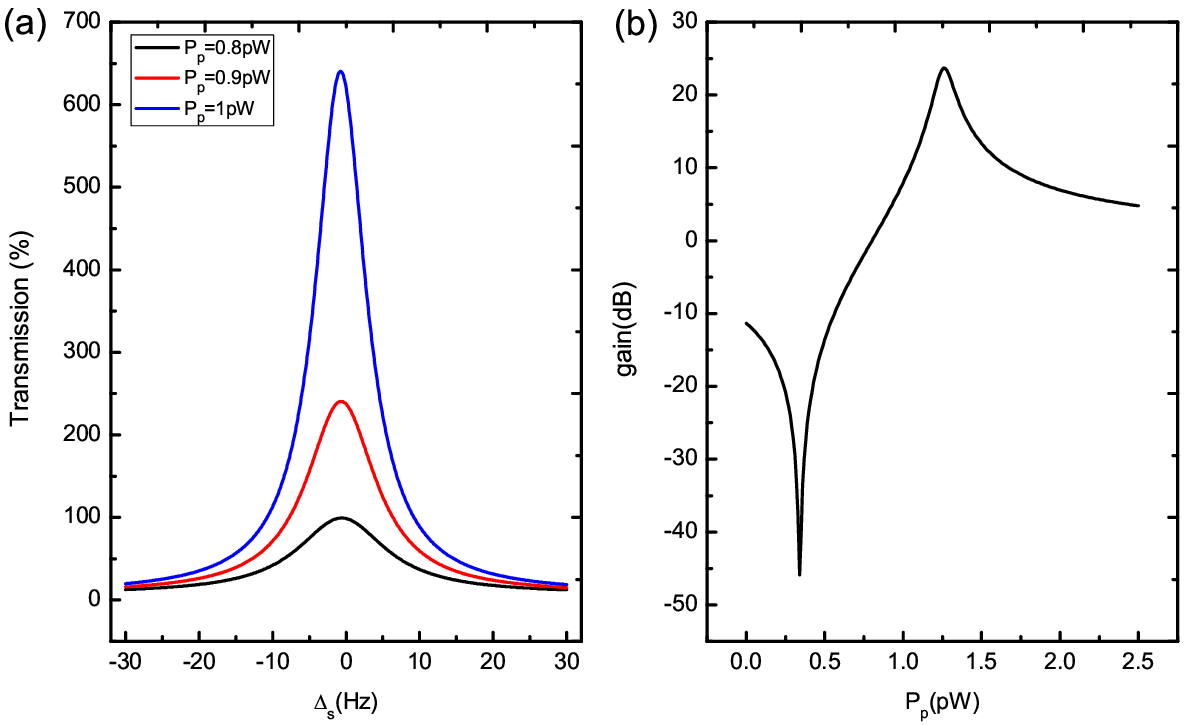}
\caption{(a) Amplification of the signal field around the region
$\Delta_s=0$ for three different pump powers with $\Delta_p=-\omega_n$. (b) The photonic
transistor characteristic curve by plotting the gain of the transmitted
signal field with respect to the pump power when the signal field
is resonant with the cavity resonance frequency.
}
\end{figure}

\section*{References}
\begin{thebibliography}{28}

\bibitem{Gibbs} H. M. Gibbs, Optical Bistability: Congrolling light
with light (Academic, Orlando, 1985).

\bibitem{Brien} J. L. O¡¯Brien, Science \textbf{318}, 1567 (2007).

\bibitem{Bouwmeester} D. Bouwmeester, A. Ekert, and A. Zeilinger, The
Physics of Quantum Information (Springer, Berlin, 2000).

\bibitem{Chang} D. E. Chang, A. S. S{\o}rensen, E. A. Demler, and M. D. Lukin, Nat. Phys. \textbf{3}, 807
(2007).

\bibitem{Hong} F. Y. Hong and S. J. Xiong, Phys. Rev. A \textbf{78}, 013812
(2008).

\bibitem{Hwang} J. Hwang, M. Pototschnig, R. Lettow, G. Zumofen, A. Renn, S. G\"{o}tzinger, and V.
Sandoghdar, Nature \textbf{460}, 76 (2009).

\bibitem{Tominaga} J. Tominaga, C. Mihalcea, D. B¨¹chel, H. Fukuda, T. Nakano, N. Atoda, H. Fuji, and T.
Kikukawa, Appl. Phys. Lett. \textbf{78}, 2417 (2001).

\bibitem{Medhekar} S. Medhekar and R. K. Sarkar, Opt. Lett. \textbf{30}, 887
(2005).

\bibitem{Huang} Y. Y. Huang and S. T. Ho, Opt. Express \textbf{16},
16806 (2008).

\bibitem{Weis} S. Weis, R. Rivi\`{e}re , S. Del\'{e}glise, E. Gavartin ,
O. Arcizet, A. Schliesser, and T. J. Kippenberg, Science
\textbf{330}, 1520 (2010).

\bibitem{Agarwal} G. S. Agarwal and S. Huang, Phys. Rev.
A \textbf{81}, 041803 (2010).

\bibitem{Safavi} A. H. Safavi-Naeini, T. P. Mayer Alegre, J. Chan, M. Eichenfield, M. Winger, Q. Lin, J. T.Hill, D.
E. Chang, and O. Painter, Nature \textbf{472}, 69 (2011).


\bibitem{Regal} C. A. Regal, J. D. Teufel, and K. W. Lehnert, Nat. Phys. \textbf{4}, 555 (2008).

\bibitem{Vitali} D. Vitali, P. Tombesi, M. J. Woolley, A. C. Doherty, and
G. J. Milburn, Phys. Rev. A \textbf{76}, 042336 (2007).

\bibitem{Woolley} M. J. Woolley, A. C. Doherty, and G. J. Milburn and
K. C. Schwab, Phys. Rev. A \textbf{78}, 062303 (2008).

\bibitem{Rocheleau} T. Rocheleau, T. Ndukum, C. Macklin, J. B. Hertzberg, A. A. Clerk, and K. C. Schwab, Nature (London) \textbf{463}, 72 (2010).

\bibitem{Teufel} J. D. Teufel, D. Li, M. S. Allman,
K. Cicak, A. J. Sirois, J. D. Whittaker, and R. W. Simmonds, Nature
\textbf{471}, 204 (2011).

\bibitem{Dobrindt}J. M. Dobrindt, I. Wilson-Rae, and T. J. Kippenberg, Phys. Rev. Lett. \textbf{101}, 263602 (2008).

\bibitem{Genes}C. Genes, D. Vitali, P. Tombesi, S. Gigan, and M. Aspelmeyer, Phys. Rev. A \textbf{77}, 033804 (2008).

\bibitem{Giovannetti}V. Giovannetti and D. Vitali, Phys. Rev. A \textbf{63}, 023812 (2001).

\bibitem{Marquardt}F. Marquardt, J. P. Chen, A. A. Clerk, and S. M. Girvin, Phys. Rev. Lett. \textbf{99}, 093902 (2007).

\bibitem{Groblacher}S. Gr\"{o}blacher, K. Hammerer, M. R. Vanner, and M. Aspelmeyer, Nature \textbf{460}, 724 (2009).

\bibitem{Boyd} R. W. Boyd  {\it Nonlinear Optics} (San Diego, CA:
Academic) (2008).

\bibitem{Kippenberg1} T. J. Kippenberg and K. J. Vahala, Opt. Express. \textbf{15}, 17172 (2007).

\bibitem{Gupta} S. Gupta, K. L. Moore, K. W. Murch, and D. M.
Stamper-Kurn, Phys. Rev. Lett. \textbf{99}, 213601 (2007).

\bibitem{Kanamoto} R. Kanamoto and P. Meystre, Phys. Rev. Lett. \textbf{104},
063601 (2010).

\bibitem{Gardiner} C. W. Gardiner and P. Zoller {\it Quantum Noise} (Springer) (2004).

\bibitem{Lezama} A. Lezama, S. Barreiro, and A. M. Akulshin, Phys.
Rev. A \textbf{59}, 4732 (1999).

\bibitem{Mollow} B. R. Mollow, R. J. Glauber, Phys. Rev. \textbf{160}, 1076 (1967).

\bibitem{Verlot} P. Verlot, A. Tavernarakis, T. Briant, P.-F. Cohadon, and A.
Heidmann, Phys. Rev. Lett. \textbf{104}, 133602 (2010).

\bibitem{Braginsky} V. B. Braginsky, S. E. Strigin, and S. P. Vyatchanin, Phys. Lett. A \textbf{287}, 331
(2001).

\bibitem{Kippenberg} T. J. Kippenberg, H. Rokhsari, T. Carmon, A. Scherer, and K. J.
Vahala, Phys. Rev. Lett. \textbf{95}, 033901 (2005).

\bibitem{Massel} F. Massel, T. T. Heikkil\"{a}, J.-M. Pirkkalainen, S. U. Cho, H. Saloniemi, P. Hakonen, M. A. Sillanp\"{a}\"{a}, Nature \textbf{480}, 351-154 (2011).

\end{thebibliography}
\end{document}